\documentclass[preprint,review,12pt]{elsarticle}

\usepackage{amssymb}

\usepackage{lineno}

\begin{document}
\begin{frontmatter}
\linenumbers

\title{The surface detector array of the Telescope Array experiment}

\author[1]{T.~Abu-Zayyad}
\author[2]{R.~Aida}
\author[1]{M.~Allen} 
\author[1]{R.~Anderson} 
\author[3]{R.~Azuma}
\author[1]{E.~Barcikowski}
\author[1]{J.~W.~Belz}
\author[1]{D.~R.~Bergman}
\author[1]{S.~A.~Blake}
\author[1]{R.~Cady}
\author[4]{B.~G.~Cheon}
\author[5]{J.~Chiba}
\author[6]{M.~Chikawa}
\author[4]{E.~J.~Cho}
\author[7]{W.~R.~Cho}
\author[8]{H.~Fujii}
\author[9]{T.~Fujii}
\author[3]{T.~Fukuda}
\author[10,20]{M.~Fukushima}
\author[11]{D.~Gorbunov}
\author[1]{W.~Hanlon}
\author[3]{K.~Hayashi}
\author[9]{Y.~Hayashi}
\author[12]{N.~Hayashida}
\author[12]{K.~Hibino}
\author[10]{K.~Hiyama}
\author[2]{K.~Honda}
\author[3]{T.~Iguchi}
\author[10]{D.~Ikeda}
\author[2]{K.~Ikuta}
\author[13]{N.~Inoue}
\author[2]{T.~Ishii}
\author[3]{R.~Ishimori}
\author[1,14]{D.~Ivanov}
\author[2]{S.~Iwamoto}
\author[1]{C.~C.~H.~Jui}
\author[15]{K.~Kadota}
\author[3]{F.~Kakimoto}
\author[11]{O.~Kalashev}
\author[2]{T.~Kanbe}
\author[16]{K.~Kasahara}
\author[17]{H.~Kawai}
\author[9]{S.~Kawakami}
\author[13]{S.~Kawana}
\author[10]{E.~Kido}
\author[4]{H.~B.~Kim}
\author[7]{H.~K.~Kim}
\author[4]{J.~H.~Kim}
\author[18]{J.~H.~Kim}
\author[6]{K.~Kitamoto}
\author[5]{K.~Kobayashi}
\author[3]{Y.~Kobayashi}
\author[10]{Y.~Kondo}
\author[9]{K.~Kuramoto}
\author[11]{V.~Kuzmin}
\author[7]{Y.~J.~Kwon}
\author[19]{S.~I.~Lim}
\author[3]{S.~Machida}
\author[20]{K.~Martens}
\author[1]{J.~Martineau}
\author[8]{T.~Matsuda}
\author[3]{T.~Matsuura}
\author[9]{T.~Matsuyama}
\author[1]{J.~N.~Matthews}
\author[1]{I.~Myers}
\author[9]{M.~Minamino}
\author[5]{K.~Miyata}
\author[9]{H.~Miyauchi}
\author[3]{Y.~Murano}
\author[21]{T.~Nakamura}
\author[19]{S.~W.~Nam}
\author[10]{T.~Nonaka\corref{cor1}}
\cortext[cor1]{Corresponding author}
\ead{nonaka@icrr.u-tokyo.ac.jp}
\author[9]{S.~Ogio}
\author[10]{M.~Ohnishi}
\author[10]{H.~Ohoka}
\author[10]{K.~Oki}
\author[2]{D.~Oku}
\author[9]{T.~Okuda}
\author[9]{A.~Oshima}
\author[16]{S.~Ozawa}
\author[19]{I.~H.~Park}
\author[22]{M.~S.~Pshirkov}
\author[1]{D.~Rodriguez}
\author[18]{S.~Y.~Roh}
\author[11]{G.~Rubtsov}
\author[18]{D.~Ryu}
\author[10]{H.~Sagawa}
\author[9]{N.~Sakurai}
\author[1]{A.~L.~Sampson}
\author[14]{L.~M.~Scott}
\author[1]{P.~D.~Shah}
\author[2]{F.~Shibata}
\author[10]{T.~Shibata}
\author[10]{H.~Shimodaira}
\author[4]{B.~K.~Shin}
\author[7]{J.~I.~Shin}
\author[13]{T.~Shirahama}
\author[1]{J.~D.~Smith}
\author[1]{P.~Sokolsky}
\author[1]{T.~J.~Sonley}
\author[1]{R.~W.~Springer}
\author[1]{B.~T.~Stokes}
\author[1,14]{S.~R.~Stratton}
\author[1]{T.~A.~Stroman}
\author[8]{S.~Suzuki}
\author[10]{Y.~Takahashi}
\author[10]{M.~Takeda}
\author[23]{A.~Taketa}
\author[10]{M.~Takita}
\author[10]{Y.~Tameda}
\author[9]{H.~Tanaka}
\author[24]{K.~Tanaka}
\author[8]{M.~Tanaka}
\author[1]{S.~B.~Thomas}
\author[1]{G.~B.~Thomson}
\author[11,22]{P.~Tinyakov}
\author[11]{I.~Tkachev}
\author[3]{H.~Tokuno}
\author[2]{T.~Tomida}
\author[11]{S.~Troitsky}
\author[3]{Y.~Tsunesada}
\author[3]{K.~Tsutsumi}
\author[2]{Y.~Tsuyuguchi}
\author[25]{Y.~Uchihori}
\author[12]{S.~Udo}
\author[2]{H.~Ukai}
\author[1]{G.~Vasiloff}
\author[13]{Y.~Wada}
\author[1]{T.~Wong}
\author[1]{M.~Wood}
\author[10]{Y.~Yamakawa}
\author[8]{H.~Yamaoka}
\author[9]{K.~Yamazaki}
\author[19]{J.~Yang}
\author[17]{S.~Yoshida}
\author[26]{H.~Yoshii}
\author[1]{R.~Zollinger}
\author[1]{Z.~Zundel}

\address[1]{University of Utah, High Energy Astrophysics Institute, Salt Lake City, Utah, USA}
\address[2]{University of Yamanashi, Interdisciplinary Graduate School of Medicine and Engineering, Kofu, Yamanashi, Japan}
\address[3]{Tokyo Institute of Technology, Meguro, Tokyo, Japan}
\address[4]{Hanyang University, Seongdong-gu, Seoul, Korea}
\address[5]{Tokyo University of Science, Noda, Chiba, Japan}
\address[6]{Kinki University, Higashi Osaka, Osaka, Japan}
\address[7]{Yonsei University, Seodaemun-gu, Seoul, Korea}
\address[8]{Institute of Particle and Nuclear Studies, KEK, Tsukuba, Ibaraki, Japan}
\address[9]{Osaka City University, Osaka, Osaka, Japan}
\address[10]{Institute for Cosmic Ray Research, University of Tokyo, Kashiwa, Chiba, Japan}
\address[11]{Institute for Nuclear Research of the Russian Academy of Sciences, Moscow, Russia}
\address[12]{Kanagawa University, Yokohama, Kanagawa, Japan}
\address[13]{Saitama University, Saitama, Saitama, Japan}
\address[14]{Rutgers University, Piscataway, USA}
\address[15]{Tokyo City University, Setagaya-ku, Tokyo, Japan}
\address[16]{Waseda University, Advanced Research Institute for Science and Engineering, Shinjuku-ku, Tokyo, Japan}
\address[17]{Chiba University, Chiba, Chiba, Japan}
\address[18]{Chungnam National University, Yuseong-gu, Daejeon, Korea}
\address[19]{Ewha Womans University, Seodaaemun-gu, Seoul, Korea}
\address[20]{University of Tokyo, Institute for the Physics and Mathematics of the Universe, Kashiwa, Chiba, Japan}
\address[21]{Kochi University, Kochi, Kochi, Japan}
\address[22]{University Libre de Bruxelles, Brussels, Belgium}
\address[23]{Earthquake Research Institute, University of Tokyo, Bunkyo-ku, Tokyo, Japan}
\address[24]{Hiroshima City University, Hiroshima, Hiroshima, Japan}
\address[25]{National Institute of Radiological Science, Chiba, Chiba, Japan}
\address[26]{Ehime University, Matsuyama, Ehime, Japan}


\begin{abstract}
The Telescope Array (TA) experiment, located in the western desert of Utah,
USA, is designed for observation of extensive air showers from extremely
high energy cosmic rays. The experiment has a surface detector array
surrounded by three fluorescence detectors to enable simultaneous
detection of shower particles at ground level and fluorescence photons along
the shower track. The TA surface detectors and fluorescence detectors started
full hybrid observation in March, 2008. In this article we describe the design
and technical features of the TA surface detector.
\end{abstract}


\end{frontmatter}
\linenumbers


\section{Introduction}
\label{Intro}
   The main aim of the Telescope Array (TA) experiment \cite{TA} is to explore
the origin of ultra high energy cosmic rays (UHECR) using
their energy spectrum, composition and anisotropy.
   There are two major methods of observation for detecting cosmic rays
in the energy region above 10$^{17.5}$ eV. One method which was used at 
the High Resolution Fly's Eye (HiRes) experiment is to detect air 
fluorescence light along air shower track using fluorescence detectors.
The other method, adopted by the AGASA experiment, is to detect
air shower particles at ground level using surface detectors 
deployed over a wide area ($\sim$100 km$^{2}$).


The AGASA experiment reported that there were 11 events above 10$^{20}$
eV in the energy spectrum \cite{AGASAGZK1998,AGASAENERGY_SYSTEMATICS2003}. 
However, the existence of the GZK cutoff \cite{GZKRONBUN,ZKRONBUN} was reported by the HiRes experiment \cite{HIRES2004}.
The Pierre Auger experiment confirmed the suppression on 
the cosmic ray flux at energy above 4$\times$10$^{19}$ eV \cite{AUGERGZK2008}
using an energy scale obtained by fluorescence light telescopes (FD).
The contradiction between results from fluorescence detectors 
and those from surface detector arrays (SD) remains to be investigated 
by having independent energy scales using both techniques.
Hybrid observations with SD and FD enable us to compare both energy scales. 
Information about core location and impact timing from SD
observation improves accuracy of reconstruction 
of FD observations. Observations with surface detectors 
have a nearly 100\% duty cycle, which is an advantage especially for 
studies of anisotropy. Correlations between arrival directions of cosmic rays and 
astronomical objects in this energy region should give a key to 
exploring the origin of UHECR \cite{AGASA18eVANISO} and their propagation 
in the galactic magnetic field. 

In this article we describe the design and technical features of the TA surface detector.


\section{Telescope Array experiment}
   The TA site is located in the desert about 1400 m above sea level
 centered at $39.3^{\circ}$N and $112.9^{\circ}$W 
in Millard County, Utah, USA, about 200 km southwest of Salt Lake City.  
A control center to support construction and operation of the TA
instruments is in the city of Delta located near the northeast side of the array. The
experiment is aimed at observing cascade showers induced by cosmic
rays above 10$^{19}$ eV. The altitude of the experimental site 
is optimal for observing particle showers at nearly maximum 
development of the cascade. For hybrid observation the
site also needed to be located in a semi-desert area with less light pollution 
from the town. 
The dry climate allows us to have a high duty cycle for FD-SD hybrid
 exposure; about 10\% of real time.

  \begin{figure}
   \begin{center}
    \includegraphics[clip,width=0.8\textwidth]{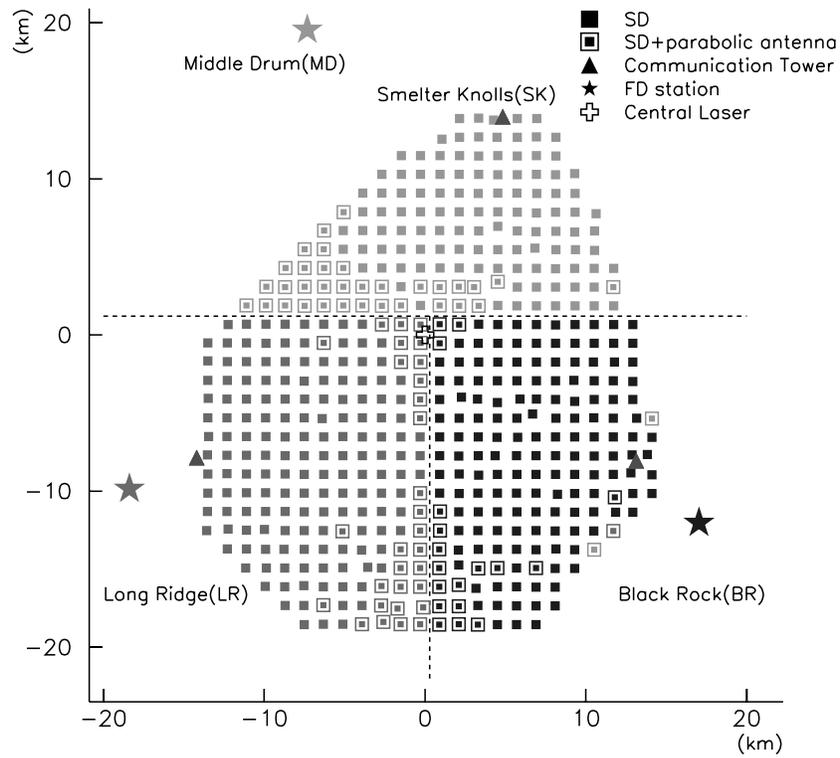}
   \end{center}
   \caption{Layout of the Telescope Array in Utah, USA. Squares denote
   507 SDs. There are three subarrays controlled by three communication
   towers denoted by triangles. The three star symbols denote the
   FD stations.}
   \label{taarray}
  \end{figure}
   Below we describe the major advantages of the TA experiment:\\
   1) The TA experiment utilizes plastic scintillators similar to AGASA
   experiment. For energies of about 10$^{20}$ eV, more than 90\%
of the primary energy is absorbed as the electromagnetic component ($e^{+}$,$e^{-}$
and $\gamma$) in the air. Plastic scintillators are sensitive to all charged
   particles, and the energy measurement is less affected by the difference of the details of
unknown hadron interactions and the primary composition.
2) The HiRes-I telescope system was partially moved to the Middle Drum
   (MD) hill in TA site and installed as one of the three FD stations 
after the HiRes experiment was shut down in 2006 \cite{MD}. 
Using an energy spectrum obtained with MD station data,
it is possible to cross-check the new TA FD data and analysis method.
 The surface detector observes lateral distribution of the shower particle. 
Energy deposition at a certain distance from shower core is used as an 
estimator of the energy by comparing with air shower Monte Carlo 
simulation. It is possible to compare the estimated energy with that 
obtained from longitudinal shower development observed by FD data analysis.\\
3) In addition to the conventional calibration and monitor system, we
plan to perform absolute end-to-end calibration of a fluorescence telescope
by using pseudo air shower events that are induced by electron beams with
known total energy from a compact electron linear accelerator at the TA
   site \cite{TA-LINAC,TA-LINAC_KEK}.
  As described above, the TA experiment is well-balanced to 
determine the energy of air shower events. \\
4) The anisotropy of arrival directions of ultra-high energy cosmic 
rays is being studied in the northern hemisphere where the effect of 
the galactic magnetic field is smaller than that in the southern 
hemisphere. A typical angular resolution of TA SD array is 
better than 1.5$^{\circ}$ for the shower above 10 EeV \cite{ICRC2011TKCHEV}.

\section{Surface detector array}
\label{SDarray}
   The SD array consists of 507 detector units, which were deployed in a
square grid with 1.2 km spacing to cover a total area of approximately
700 km$^{2}$.  Fig.~\ref{taarray} shows a layout of the TA
   experiment. Each surface detector has a plastic
scintillation counter of 3 m$^{2}$ in size, and transmits SD data via 
a wireless LAN modem. As shown in Fig.~\ref{taarray}, the SD array is divided into 
three subarrays each controlled by its trigger-decision electronics at 
the communication tower. The Long Ridge (LR), Black Rock (BR),
and Smelter Knolls (SK) subarrays have 189, 170 and 148 SDs respectively. 
(The numbers of SDs in LR , BR and SK from March 2008 to November 2009 were 207, 190
and 110 respectively.) 
All detectors are powered by solar panels and batteries \cite{TOMIDA_SYUURON}.
  For events with energies beyond 10$^{19}$ eV and with zenith angles 
below 45 degrees, the trigger efficiency reaches $\sim$100\% and the aperture
 is 1100 km$^{2}$$\cdot$sr.
The observed energy region for the TA experiment has sufficient overlap
   with those for the previous experiments of UHECR.

%
\subsection{Surface detector}
   The TA detector will operate for more than 10 years and must be 
designed to survive the expected conditions at the site.
The detector must be robust and durable for long-term exposure 
to the desert environment where the detector 
temperature ranges from $-20^{\circ}$C to $+50^{\circ}$C with large
diurnal variations. And the system requires detailed monitoring and
periodic calibrations to track variations of detector response along
time.
  \begin{figure}
   \begin{center}
    \includegraphics[clip,width=0.8\textwidth]{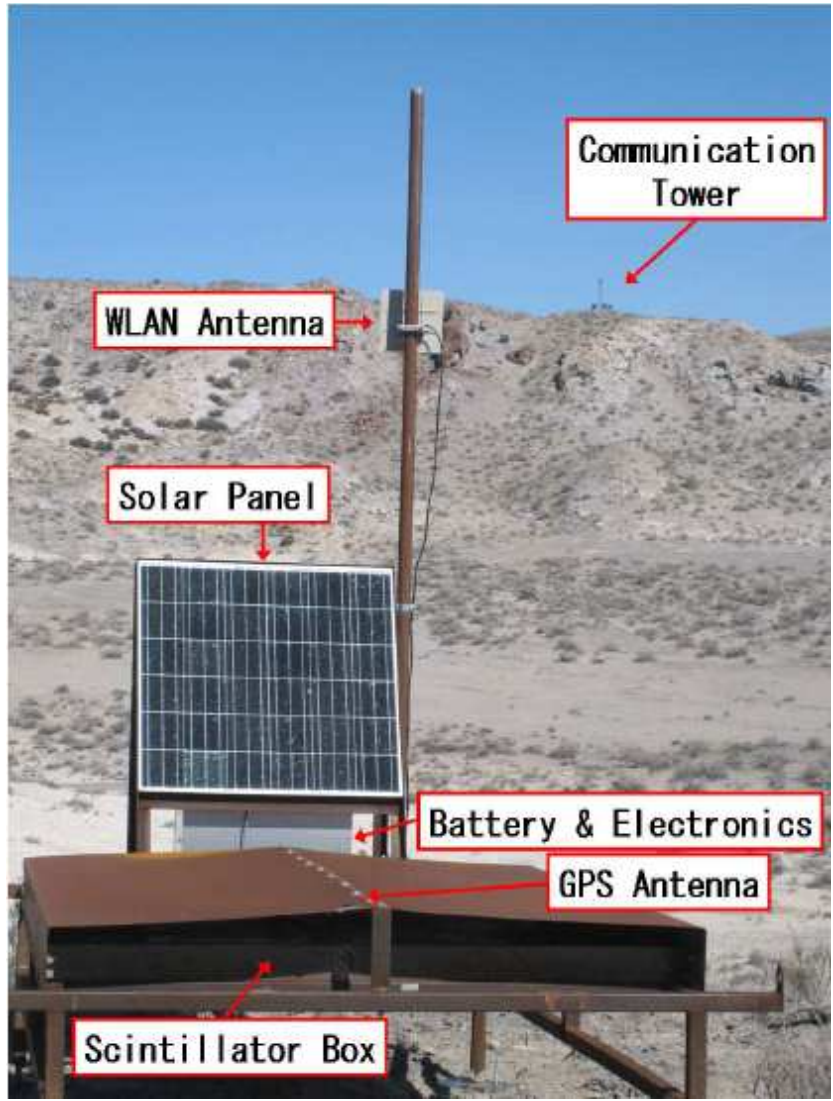}
   \end{center}
   \caption{A deployed SD in the field. The electronics box and
   scintillator box are on the iron frame. An electronics unit is installed
   under the solar panel, and the scintillator box is mounted on the platform
   under the roof.}
   \label{sdsetumei}
  \end{figure}

   Fig.~\ref{sdsetumei} shows one of the deployed SDs that communicate with the 
communication tower placed at Smelter Knolls (SK), a nearby hill. 
A communication antenna (ADAF2414; ADTEC Co.) with adjustable height is mounted on
a 3-m long iron pole. A square solar panel 1 m on one side is mounted on
the platform to supply power to the electronics. Front-end electronics and a
battery are contained in a box made with 1.2 mm thick stainless steel.
The box is mounted under the solar panel. The box that contains the scintillators and photomultiplier
tubes (PMTs) is mounted under the 1.2-mm thick iron roof to protect 
the detector from large temperature variations.
  \begin{figure}
   \begin{center}
    \includegraphics[clip,angle=270,width=0.8\textwidth]{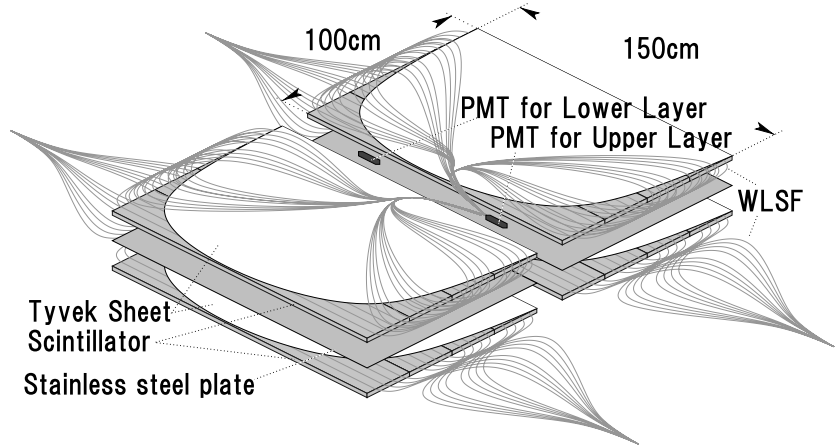}
   \end{center}
   \caption{Inside of a scintillator box with scintillator plates,
   WLS fibers and PMTs. A total of 104 WLS fibers are laid on each layer to
   collect and transmit scintillation light to a PMT \cite{S.KAWAKAMI2005}.}
   \label{fiber}
  \end{figure}

   Fig.~\ref{fiber} shows a schematic of the inside of a scintillator box. 
Each surface detector consists of two layers of plastic scintillator. 
Each layer of scintillator has an area of 3~m$^{2}$ and a thickness of
1.2 cm. A Stainless-steel plate has 1 mm in thickness and is inserted 
between the layers. 
As shown in Fig.~\ref{fiber}, each scintillator layer consists of 2
 segments, and each segment consists of 4 slabs. 
The size of one segment is 1.5 m $\times$ 1.0 m. 
The size of each slab is 1.5 m $\times$ 0.25 m and thickness is 1.2 cm.
On top side of the scintillator slab, there are grooves in parallel 
along the length of the slab. The span of grooves is 2.0 cm and the depth
is 1.5 mm. Scintillation light is collected through 104 wavelength-shifting (WLS)
 fibers (Y-11; Kuraray Co. Ltd.) that are laid along each groove.
Total length of a WLS fiber is 5 m. 
The fibers are put in the grooves on the surfaces of the scintillator
slabs without oil and grease, and are fixed at both edges of the slabs
with tape (polyester tape \#850 silver; 3M). 
The segment is wrapped with two layer of Tyvek (1073B; Dupont Co.) sheet. 
Both ends of the fibers from a layer are bundled together and connected to 
a PMT (9124SA; Electron Tubes Ltd.).

Each PMT is calibrated to obtain the relation of high voltage and gain. 
Linearity between input light amount and output charge is also obtained
in the calibration \cite{S.YOSHIDA2005}. Two LEDs (NSPB320BS; Nichia
Corp.) are also installed on the side of each layer to calibrate 
linearity of output for input light.
Scintillator plates and PMTs are contained in a 1.5 mm thick box made of
stainless steel (top cover is 1.5 mm thick, with a 1.2 mm thick bottom) (TAITO
Co. Ltd.).

  \subsection{Detector Electronics}
  \begin{figure}
   \begin{center}
    \includegraphics[clip,width=1.0\textwidth]{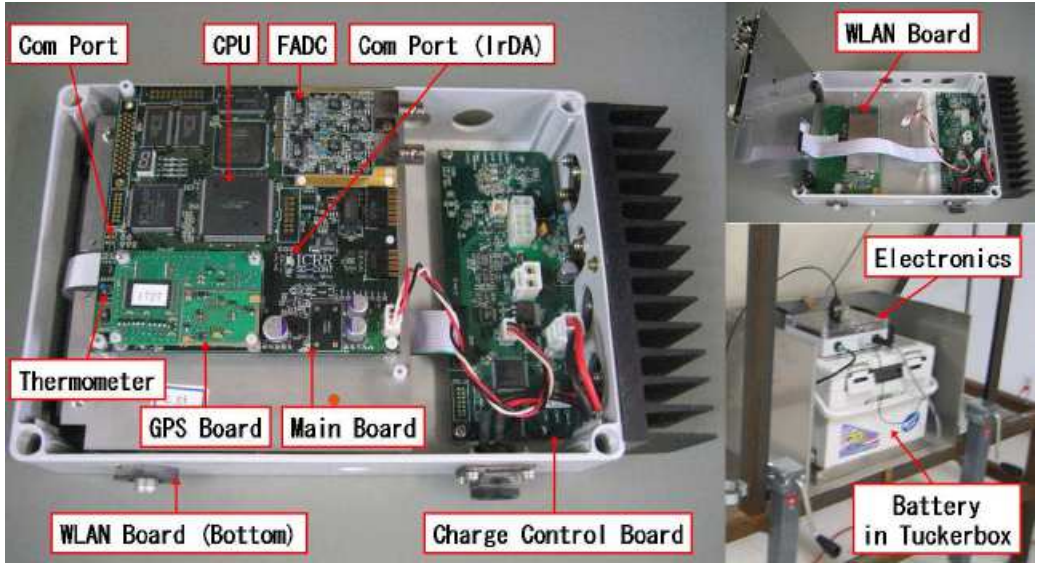}
   \end{center}
   \caption{Detector electronics of the TA surface detector. The wireless
   LAN board is mounted under the main CPU board.}
   \label{sdelec}
  \end{figure}
  Fig.~\ref{sdelec} shows the detector electronics for a scintillator counter 
installed in a stainless-steel box under the solar panel. 
  The output signals from PMTs are digitized by a 12bit FADC (AD9235RU-
65; Analog Devices Co.) with a 50 MHz sampling rate on the CPU board
(SH4; Renesas Electronics Co.). Signals greater than approximately 0.3 minimum ionizing particles (MIP) are stored
in a memory buffer on CPU board as Level-0 trigger data. The stored waveform is 2.56~$\mu$sec long (128 FADC bins). Signals greater than 3.0 MIP are stored as a Level-1
trigger event, which are sent to the trigger-decision electronics at 
the communication tower for the subarray via a wireless LAN
modem (ADLINK540F; ADTEC. Co) using a custom-made communication
protocol \cite{ATAKETA2009}.
  The local trigger rates are $\sim$750 Hz for the Level-0 trigger and $\sim$30 Hz 
for the Level-1 trigger. In FPGA, the FADC pedestals are 
monitored every second to keep threshold values for the
trigger. The pedestal value is defined as the average of FADC values within
160 nsec (8 FADC bins). The threshold value for Level-0 trigger is fixed to 15 counts above 
the average of pedestal.
Triggered waveforms mostly from atmospheric particles 
are counted into histograms based on integrated count and maximum count of the waveform. 
For the monitoring of detector gain, the signal part of the 
waveform is integrated for 240 nsec (-80 nsec from trigger timing and
up to +160 nsec after) and the values are accumulated into 
histograms for each layer of scintillator.
The integrated FADC value and the maximum value of the signal waveform
within 2.56 $\mu$sec are also accumulated in histograms.
These are used for detector linearity monitoring.
The synchronization of electronics of the surface detectors is done by
PPS signals received by GPS units (Motorola M12+ oncore module). A time stamp
with a precision of 20 nsec is created by the 50 MHz sub-clock on the
main board. The total counts of the sub-clock between PPS signals are also sent
to the trigger-decision electronics along with a Level-1 event list to correct
the time stamp of the waveform in later analysis.
The power bases (PS1806/12F-02; Electrontube Co. Ltd.) for PMTs are 
powered and controlled through DAC on the detector electronics.
Each SD unit described above is powered by one
solar panel (KC125TJ; KYOCERA Corp.)  and one deep cycle battery
(DCS100; C\&D technologies, Inc.). The solar panel has 125 W of charging
power. The battery has 100 Ah of capacity. The charging of the battery is controlled
by home made charge control board that works with main CPU board.  The
solar panel system provides sufficient power required from the electronics ($\sim$5 W).

\subsection{Assembly and deployment}
  Detector assembly is divided into two parts.
First, two layers of plastic scintillator are installed in the
stainless-steel  box and WLS fibers are laid on each layer.
To increase the light intensity, each layer of scintillator is wrapped
with two layers of Dupont Tyvek sheeting. Both ends of a bundle of WLS
fibers from a layer are glued together with epoxy inserted into an
acrylic sheath. The sheath is sized to fit the PMT surface. 
The ends of the WLS fibers are smoothed with a grinder and polished 
after making a bundle in the sheath. 
To ensure good optical contact, optical grease (Optseal; Shin-Etsu Chemical
Co. Ltd.) is also applied on the surface of the PMT. 
The production rate of scintillator boxes was three boxes per day.  
In total 512 SD boxes were assembled in Japan and those boxes 
were shipped to Utah, USA.
Second, the final assembly of other components such as solar panels, batteries, and
electronics along with mounting these on to iron 
frames (T\&D Co. Ltd.) was performed at the control center in Delta. 
Each SD unit was deployed to its location by using helicopter 
after transporting the units by trucks with a flatbed trailer to staging 
areas accessible from existing roads inside the TA site. 
From October 2006 to the end of February 2007, 485 surface detectors 
were deployed. A total of 503 SDs were deployed by the end of December 2007. 
Additional 4 SDs were deployed in December 2008.

\subsection{Long-distance network for remote operation}
   \begin{figure}
    \begin{center}
     \includegraphics[clip,width=0.8\textwidth]{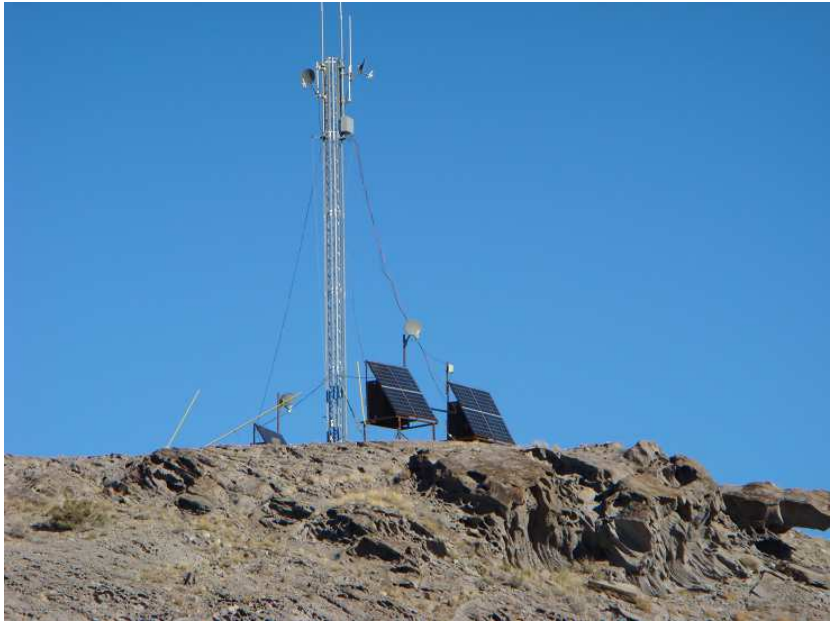}
    \end{center}
    \caption{
    The Smelter Knolls communication tower, one of three in the array.
    There are three stands each with four solar panels. Those stands contain batteries, data acquisition PC
    and network instruments for long distance link.
    }
    \label{sktower}
   \end{figure}

   Fig.~\ref{sktower} shows one of the communication towers, which is located at
Smelter Knolls (SK) near the north edge of the SD array. The other two
towers also have the same size and are located on hills (Black Rock and
Long Ridge) near the western and eastern edges of the SD
array, respectively. The communication towers have the role of collecting
trigger information from the SDs and providing communications for the FD
stations and CLF (Central Laser Facility) \cite{UDOCLF2007} site for
general purposes.
   \begin{figure}
    \begin{center}
     \includegraphics[clip,width=0.8\textwidth]{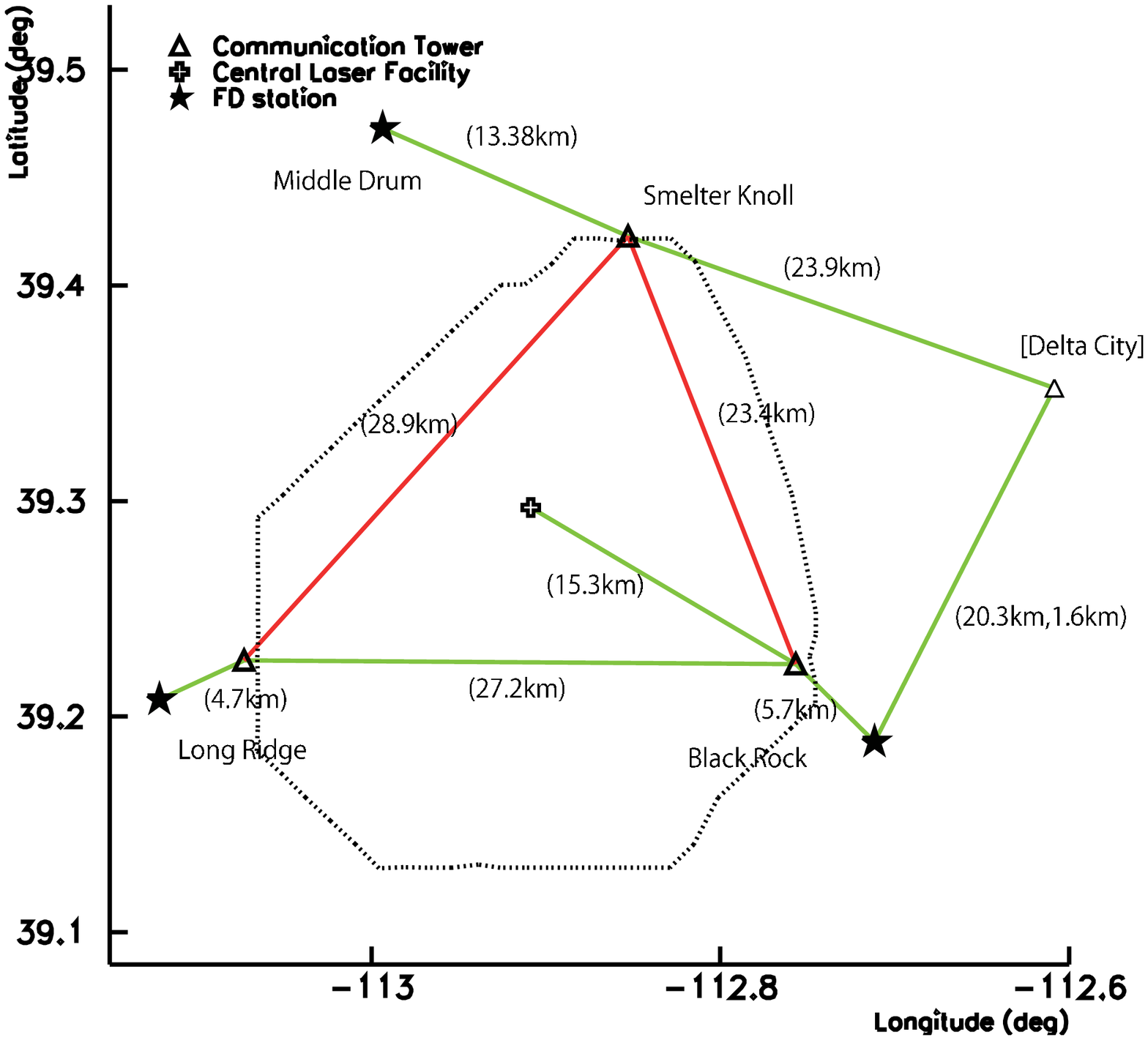}
    \end{center}
    \caption{
    The long-distance links for all the facilities and three 
    FD stations in the entire TA site. The open triangles represent 
    the communication towers where the trigger-decision electronics 
    for subarrays are installed. The lines that connect the 
    towers and facilities represent the links between antennas. The red
    lines are used for trigger decision.
    The dotted line represents the border of the entire surface detector array.}
    \label{tanetwork}
   \end{figure}

   Fig.~\ref{tanetwork} shows the long-distance links for all the facilities and air 
fluorescence detectors in the entire TA site. The open triangles in 
Fig.~\ref{tanetwork} represent
the locations of the communication towers. The data collected from the
SDs are temporarily stored at the communication tower and regularly 
transferred to Delta City through this network every 12 hours.
There are two types of antenna units (Canopy P2P100; Motorola Co. Ltd.)
which are operated in different frequency ranges. The tower-to-tower and
tower-to-FD links are operated at 5.7 GHz and the tower-to-SD links are operated at 2.4 GHz range. 
The line between the SK and
BR towers and the line between the SK and LR towers are used 
for SD data acquisition. No access to FD stations and CLF for 
general purposes interferes with the SD data acquisition lines.
   The long distance networks are currently operated at 3 Mbps 
throughput, which is sufficient for SD data acquisitions, 
data transfer and operation of FD stations.
   \begin{figure}
    \begin{center}
     \includegraphics[clip,angle=270,width=0.8\textwidth]{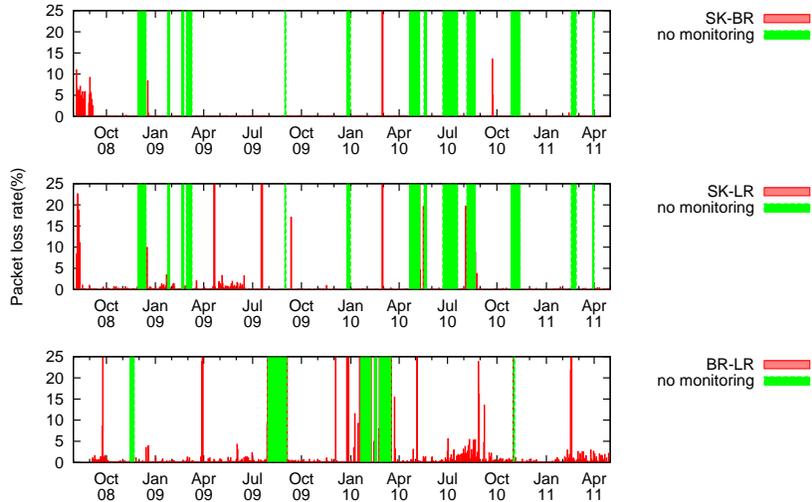}
    \end{center}
    \caption{
    Long-term stability of the network from August 2008. 
    Red histograms in the top two panels show daily packet 
    loss rate in the line between the SK to BR towers and 
    in the line between the SK to LR towers, respectively.
    The bottom panel shows the same in the line between the BR and 
    LR FD stations. The monitor data for daily packet loss rate were not
    recorded in the green regions.}
    \label{pktloss}
   \end{figure}

 Fig.~\ref{pktloss} shows the daily average values of packet loss rate in the
lines monitored in the current system. The network is stable
enough, except for some periods of major maintenances. More details of 
long distance network in the TA experiment are described in papers \cite{NONAKAICRC2009tower,IWAMOTO_SYUURON}.

\section{Data communication and air shower trigger}
There is a trigger-decision module at each tower. The electronics 
is the same as shown in Fig.~\ref{sdelec}, running a different firmware 
program. The data communication between the trigger-decision electronics at
communication tower and SDs is done by 2.4 GHz wireless LAN using a
custom-made communication protocol. The baud rate of the
data acquisition is $\sim$1 Mbps. Every second, the trigger-decision 
electronics at the communication tower requests each surface detector 
to send a Level-1 trigger event list and the total counts of the 
sub-clock between PPS signals. From the event lists, an air shower trigger
is generated when three adjacent SDs are coincident within 8~$\mu$sec.
  \begin{figure}
   \begin{center}
    \includegraphics[clip,width=0.5\textwidth]{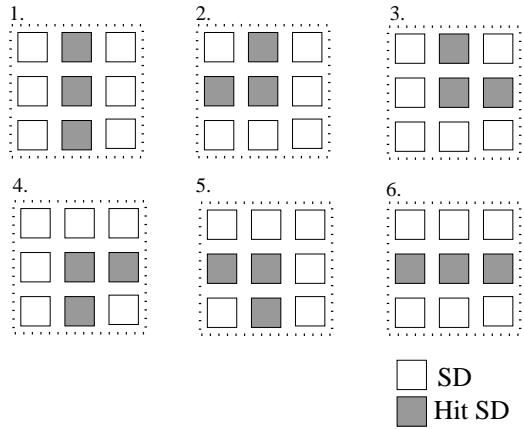}
   \end{center}
   \caption{
   Trigger pattern taken at the TA surface array. If any three 
adjacent SDs have Level-1 trigger, timing differences are within 
8~$\mu$sec of which the Level-2 trigger will be generated.}
   \label{triggerpattern}
  \end{figure}
Fig.~\ref{triggerpattern} shows the current trigger pattern of three adjacent SDs that are
hit. We call this trigger the Level-2 trigger. With this trigger,
collection of waveforms stored in each SD electronics starts. The trigger-decision electronics collects waveforms
coincident within $\pm$32~$\mu$sec from the trigger timing.
When the Level-2 trigger is generated within one subarray, the 
trigger time information is transmitted to the central trigger 
decision process. The process is running at the data acquisition 
PC (TS7800; Technologic systems, Co. Ltd.) in the SK tower.
From the SK tower, the Level-2 trigger signal is distributed to other two towers.
The broadcasting of this trigger enables to collect waveforms associated
with a shower which impacts multiple subarrays.
\begin{figure}
  \begin{center}
   \includegraphics[clip,angle=270,width=0.8\textwidth]{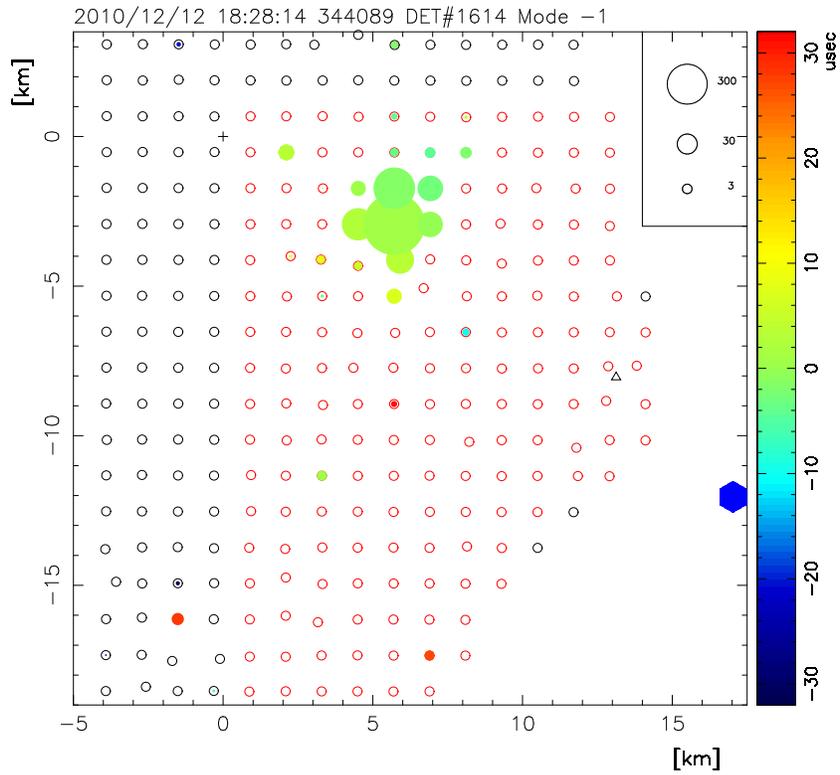}
  \end{center}
  \caption{An example of triggered event at BR array. Red open circles
  represent SDs in BR subarray. The triggered SDs are shown with color
 code, which corresponds to the arrival time. The radius of a circle is proportional to 
the logarithm of the integrated signal amount in the unit of MIP.}
  \label{eventsample}
\end{figure}
\begin{figure}
   \begin{center}
    \includegraphics[clip,angle=270,width=0.8\textwidth]{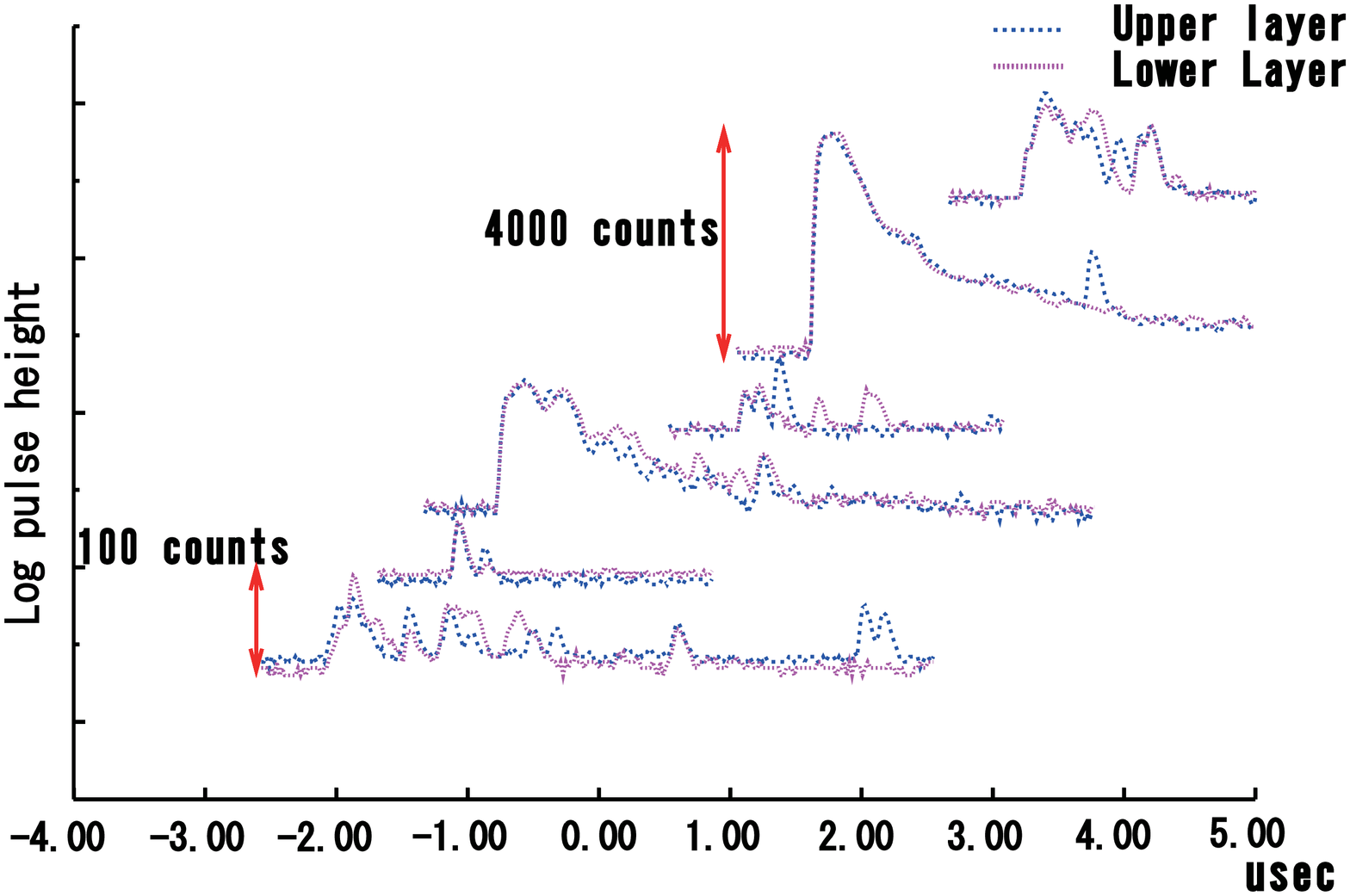}
   \end{center}
   \caption{Waveforms obtained from the event shown in Fig.~\ref{eventsample}}
   \label{wfsample}
\end{figure}
Fig.~\ref{eventsample} shows an event triggered at BR
subarray. Fig.~\ref{wfsample} shows a sample of waveforms.\\
To collect air shower event at the boundary of subarrays,
the Level-1 trigger event lists from the SDs at the boundary of subarrays are also sent 
to the central trigger process in SK tower. 
The central trigger process combines Level-1 trigger information 
collected from SDs at the boundary of subarrays.
The central trigger process  verifies whether the same trigger condition 
was satisfied only by the boundary detectors.\\
%
%
As shown in Fig.~\ref{triggerpattern}, in the case of pattern 1 or 6 ,
the hit pattern that satisfies the Level-2 trigger condition can be spread over 2
subarrays, but the condition is not satisfied with edge detectors only. 
To trigger such a case, central trigger process searches for the coincidence
of two adjacent hits in the boundary detectors from Level-1 trigger list. 
If such a coincidence exists, the central trigger process sends time and
position information of the coincidence to all towers. 
Trigger-decision electronics at towers verify the Level-2 trigger condition using
the received information from the central trigger process. Shower events of relatively small size that fall near the boundary
of subarrays is collected by this trigger scheme.

With the above trigger conditions, trigger efficiency
reaches 97\% for primary particle with energy of 10$^{19.0}$ 
eV. Here the efficiency includes the effect of dead counters~\cite{ATAKETA2009,NSAKURAIICRC2007,YAMAKAWA2009}.
   \begin{figure}
    \begin{center}
     \includegraphics[clip,width=0.7\textwidth]{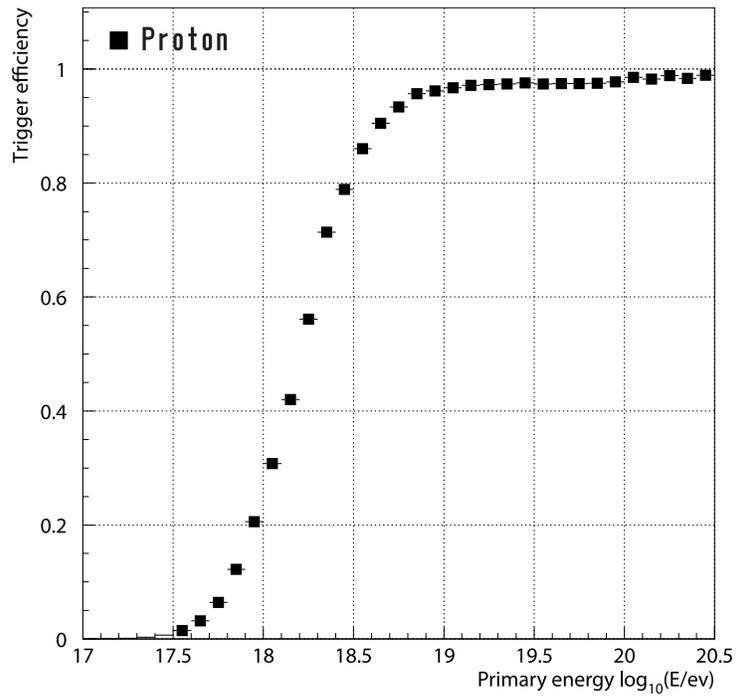}
    \end{center}
    \caption{
    Trigger efficiency as a function of energy for primary proton calculated using CORSIKA air shower Monte Carlo simulation 
    and GEANT4 detector Monte Carlo simulation.}
    \label{triggerefficiency}
   \end{figure}
   Fig.~\ref{triggerefficiency} shows trigger efficiency as a
function of energy of primary particle obtained using CORSIKA \cite{CORSIKA_HECK} 
air shower Monte Carlo simulation and GEANT4 \cite{GEANT4} 
detector Monte Carlo simulation \cite{BENSDMC}.

To improve the efficiency of detecting FD-SD hybrid events at 
lower energies ($<$ 10$^{18.7}$ eV), an external trigger called 
Hybrid Trigger was installed to collect waveforms associated 
with air showers detected by FD. From nearest FD, 
the SD data acquisition system at a tower receives trigger timing 
and time window for requesting waveforms stored at each SD 
in the subarray.
The efficiency of the waveform collection is greater than $97 \%$ 
for the primary particle with energy of 10$^{17.5}$ eV triggered by FD  
\cite{RISHIMORI_HYBRID}. The extended hybrid trigger observations 
started in October 2010.

\section{Detector Calibration and Monitoring}

   For stable observation, the status and environment of the batteries need
to be monitored continuously. For calibration in later analysis, it is very
important to monitor the detector response. For this purpose, a 
monitoring process runs on each SD in a 10 min cycle.
The monitored items are summarized in Table 1 \cite{NONAKAICRC2009}. 
The size of monitoring data for 10 min is 9600 bytes. 
The data is divided into 600 subsets. All the subsets are sent along 
with the Level-1 trigger tables within 10 min.

\begin{table*}[thbp]
 \caption{Items and resolutions of TA Surface Detector monitor.}
 \label{table_wide}
 \small 
 \centering
 \begin{tabular}{|c|c|c|}
  \hline
  Item  &Data  & Resolution \\
  \hline 
  1MIP histogram         & 12bin sum of FADC    & 1 FADC count 10min\\
  Pedestal histogram     &  8bin sum of FADC    & 1 FADC count  10min\\
  Pulse height histogram  & Maximum FADC        & 32 FADC count 10min\\
  Total charge histogram & 128bin sum of FADC   & $\Delta\log_{2}$(FADC sum)=0.2\\    
  Power cycle data      & Battery (voltage, current)     &  1min \\
  Environmental data    & Temperature, humidity          &  1min \\
  Trigger rate          & Level-0, Level-1 trigger rate  &  1min\\
  GPS status            & Number of satellites, status   & 10min \\
  \hline
 \end{tabular}
\end{table*}

\subsection{The 1-MIP monitor}
   The charge output by atmospheric charged particles is used to 
estimate the total energy deposited by shower particles.
The integrated FADC value recorded by the Level-0 trigger is collected
as monitoring data from each surface detector.
Here the time window for the integration is 240 nsec (12 bins).
The time window ranges between -4 bins from trigger timing and +8 bins after
trigger timing. That is sufficient to evaluate MIP peak count of FADC.
   \begin{figure}
    \begin{center}
     \includegraphics[clip,width=0.8\textwidth]{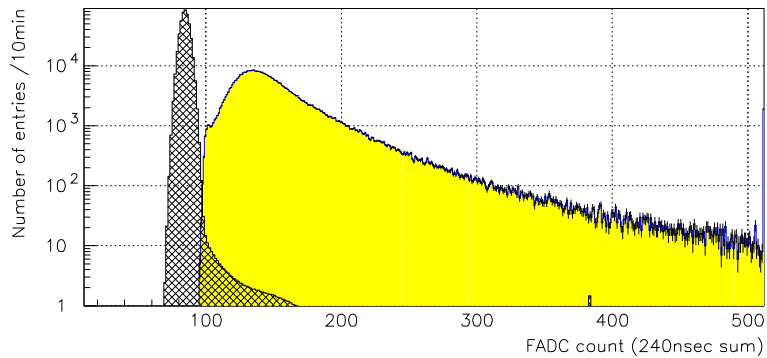}
    \end{center}
    \caption{
    Example of FADC count distribution from Level-0 trigger events
    obtained as 1-MIP
    monitor data. The hatched histogram is a pedestal distribution 
    collected as monitor data. The pedestal distribution is scaled to
    have the same entry as the Level-0 trigger data.
    The same distribution is collected from every PMT at 
    10 min intervals.}
    \label{1mipsample}
   \end{figure}
   \begin{figure}
    \begin{center}
     \includegraphics[clip,angle=270,width=0.8\textwidth]{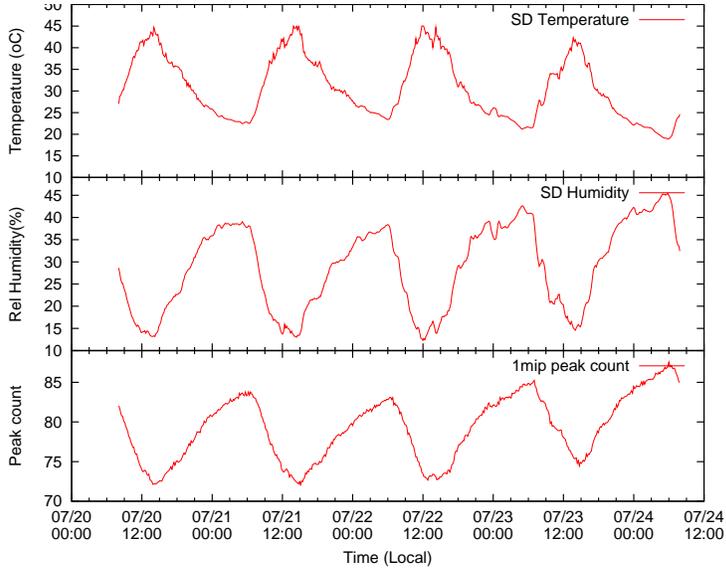}
    \end{center}
    \caption{
    Examples of time variations of temperature (top) and relative 
    humidity (middle) inside the scintillator box, and the 1-MIP peak value
    (bottom).}
    \label{1mipvariation}
   \end{figure}
Fig.~\ref{1mipsample} shows an example of the charge output distribution.
The last bin of the histogram is the overflow bin.  
The temperature coefficient of gain of the surface 
detectors is typically -0.8\%/$^{\circ}$C for a diurnal 
variation of temperature that reaches up to 25$^{\circ}$C
\cite{NONAKAICRC2007}.  Fig.~\ref{1mipvariation} shows examples of the monitored 
time variations of temperature, relative humidity inside the
scintillator box, and 1-MIP peak value.
 Change of detector response caused mainly by the variation of the
 outside temperature is monitored by this distribution continuously to the nearest ten minutes. 


\subsection{Linearity monitor}
  A check on the linearity of charge output was performed by using two LEDs
attached at each layer of scintillator. This was done for all the detectors
before deployment. Two LEDs were flashed with 400 nsec wide 
square pulses alternately or simultaneously.
The amount of light from each LED is changed by changing the pulse height 
of the square pulse.
From the ratio of the measured output to the expected output, the linearity curves 
were measured.
Here we describe the light amount as $x$ and the peak of the pulse from
a PMT in FADC count as $F(x)$ while driving one LED.  
The PMT output while driving LED 1 and 2 simultaneously is
represented as $F(x_{1}+x_{2})$.  The linearity was checked by comparing $F(x_{1}+x_{2})$ and
$F(x_{1})+F(x_{2})$ while changing $x_{1}$ and $x_{2}$.
   \begin{figure}
    \begin{center}
     \includegraphics[clip,width=0.8\textwidth]{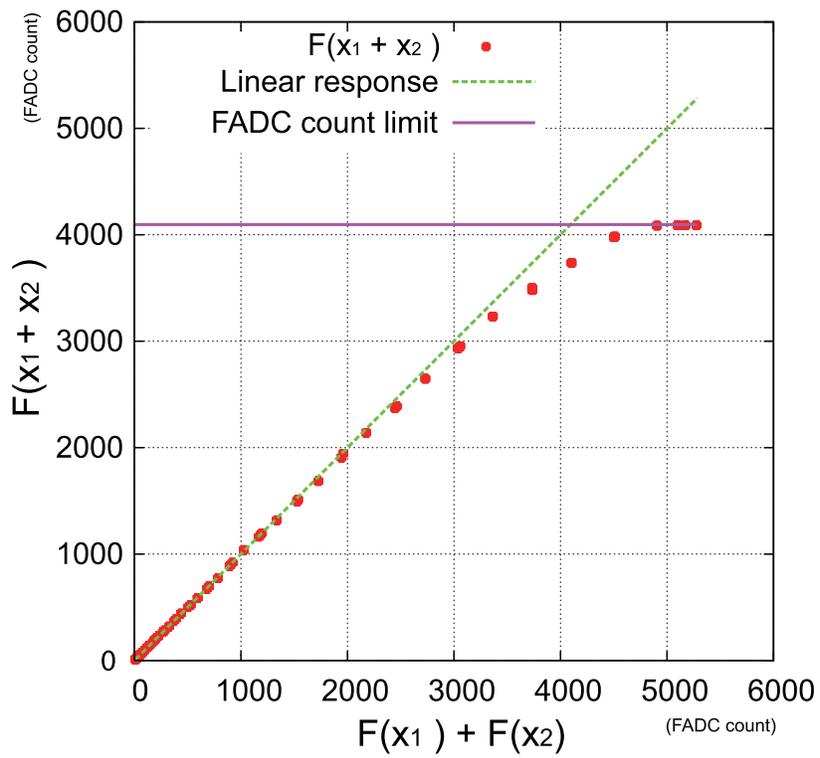}
    \end{center}
    \caption{
     An example of observed relation between $F(x_{1}+x_{2})$ and $F(x_{1})+F(x_{2})$.
    }
    \label{chlin}
   \end{figure}
Fig.~\ref{chlin} shows a typical example of observed relation between 
$F(x_{1}+x_{2})$ and $F(x_{1})+F(x_{2})$. 

 To check the linearity and its variation in the long term of
operation, pulse-height (FADC peak) histograms are also taken as
monitoring data.
The pulse height of the signal that satisfies the level-0 trigger condition is counted
into a histogram. 
The high voltage values of the PMTs of all the detectors are adjusted to
obtain almost the same FADC counts for 1 MIP peaks.
So the histograms differ between detectors because of
the difference of the linearity of the PMTs. We monitor the histogram to
detect time variation of the linearity. By comparing the tail of the
histogram and the one from the tubes with good linearity, it is 
possible to estimate non-linearity.
   \begin{figure}
    \begin{center}
     \includegraphics[clip,width=1.0\textwidth]{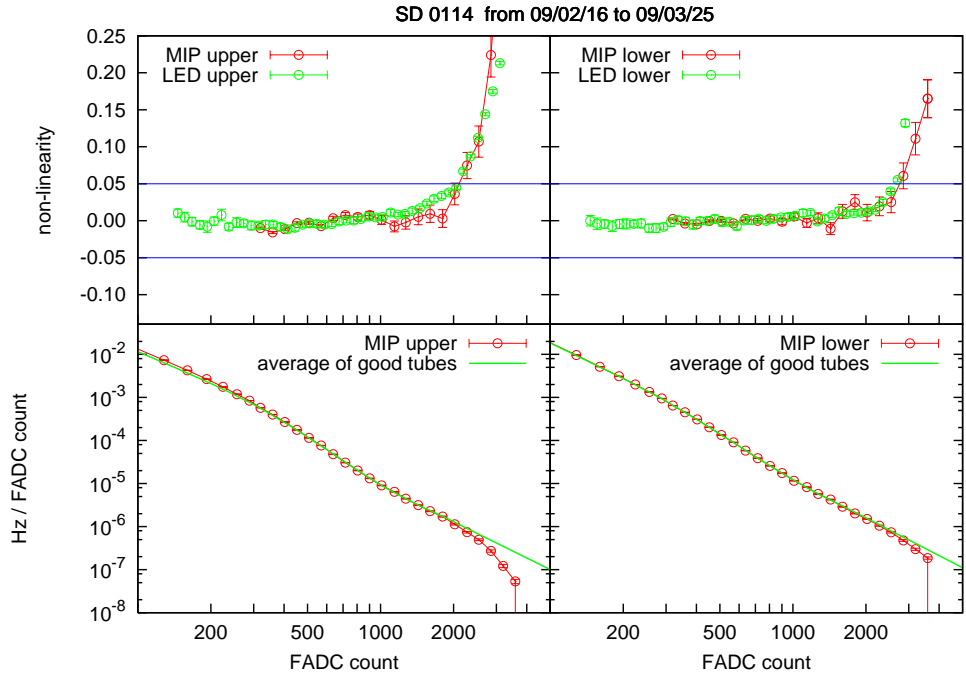}
    \end{center}
    \caption{
    Analyzed linearity of pulse height from monitor data
    \cite{ATAKETA2009}. 
    In the top left panel, the green open circles denote the 
    relative deviation from the estimated linear response by LED
    calibration, and the red open circles denote the relative 
    deviation by pulse-height monitor for the upper layer of one SD (SD \#0114). 
    The top right panel shows the same for the lower layer. 
    In the bottom left panel, the red open circles denote the scaled 
    pulse height distribution observed for the upper layer, and the
    green line denotes the average from good tubes. 
    The bottom right panel shows the same for the lower layer.
    }
    \label{phlin}
   \end{figure}
   Fig.~\ref{phlin} shows an example of comparison between 
pulse-height linearity obtained from LED calibration and
the one estimated using the pulse-height monitor.  It shows fairly good 
agreement and it was confirmed that the histogram can be used for monitoring
the stability of linearity. When the pulse height of signal is larger
than the expected height of saturation, the signal is not used for analyzing lateral
distribution of shower particle. But the timing information is used for 
calculating arrival direction. 

\subsection{Power monitor, GPS and environmental parameters}
   Since each SD is locally powered by a solar panel and a battery, it is
very important to monitor the status of the output voltage and current from the battery. The 1PPS signals are generated by GPS modules using the
signals from the satellites that are visible through the GPS antennas. To
understand the status of the GPS module, the number of satellites visible
through the GPS module and conductivity of the antenna are read out every
600 sec.
 Each surface detector is equipped with five temperature sensors and two
humidity sensors to record the environment of the detector and 
electronics box.
   \begin{figure}
    \begin{center}
     \includegraphics[clip,width=0.8\textwidth]{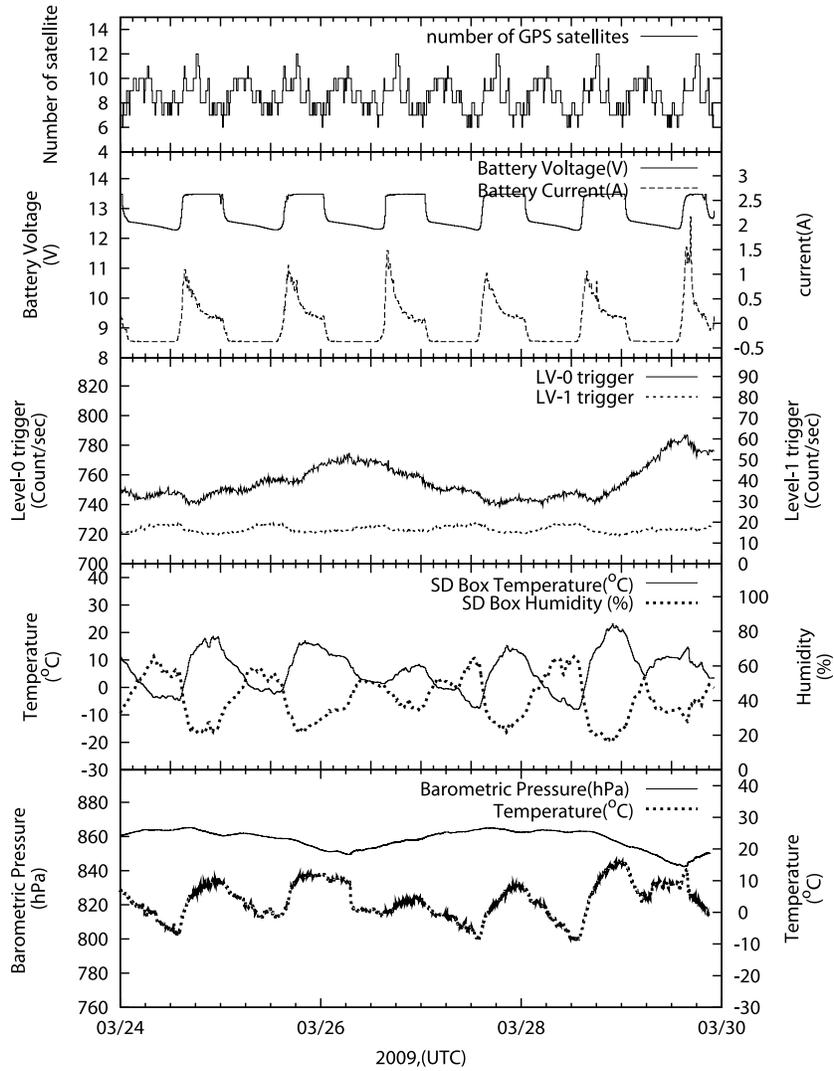}
    \end{center}
    \caption{
    An example of time variations in the number of GPS satellites 
    (top panel),
    battery voltage and charging current (second panel) , local Level-0 
    and Level-1 trigger rates (third panel) for one surface detector, 
    and the barometric pressure and atmospheric temperature measured 
    at the CLF site using a weather transmitter (WXT510; VAISALA Inc.)(bottom panel).}
    \label{monitorsample}
   \end{figure}

   Fig.~\ref{monitorsample} gives an example of the time variations 
in the number of detected GPS satellites, the battery voltage and 
charging current, the Level-0 and Level-1 trigger rates for one 
detector, and the barometric pressure and atmospheric
temperature measured at the CLF \cite{UDOCLF2007} site.

\section{Summary}
\label{conclusion}
 The SD array of the TA experiment consists of 507 plastic 
scintillation detectors of 3 m$^{2}$ in size. 
The array has the largest total area 
in the northern hemisphere. The detector enables us to compare 
estimated energy of primary particle using longitudinal 
shower development observed at FD and lateral distribution 
of shower particles detected with the SD array. The observation with 
the SD array is continuous to have $\sim$100\% of duty cycle.
This feature enables us to explore the anisotropy of arrival directions
of highest energy cosmic rays with larger exposure than observation with
 FD.


The deployment of the surface detectors started in October 2006.
Totally 507 surface detectors were deployed by November 2008.
The deployed detectors have been calibrated and tuned.
  The air shower array began operation in March 2008.
For more than three years, air shower events from UHECR 
have been collected along with detailed monitoring data.
  The monitoring data enables us to calibrate the 
variation of detector responses with enough accuracy.
   \begin{figure}
    \begin{center}
     \includegraphics[clip,angle=270,width=1.0\textwidth]{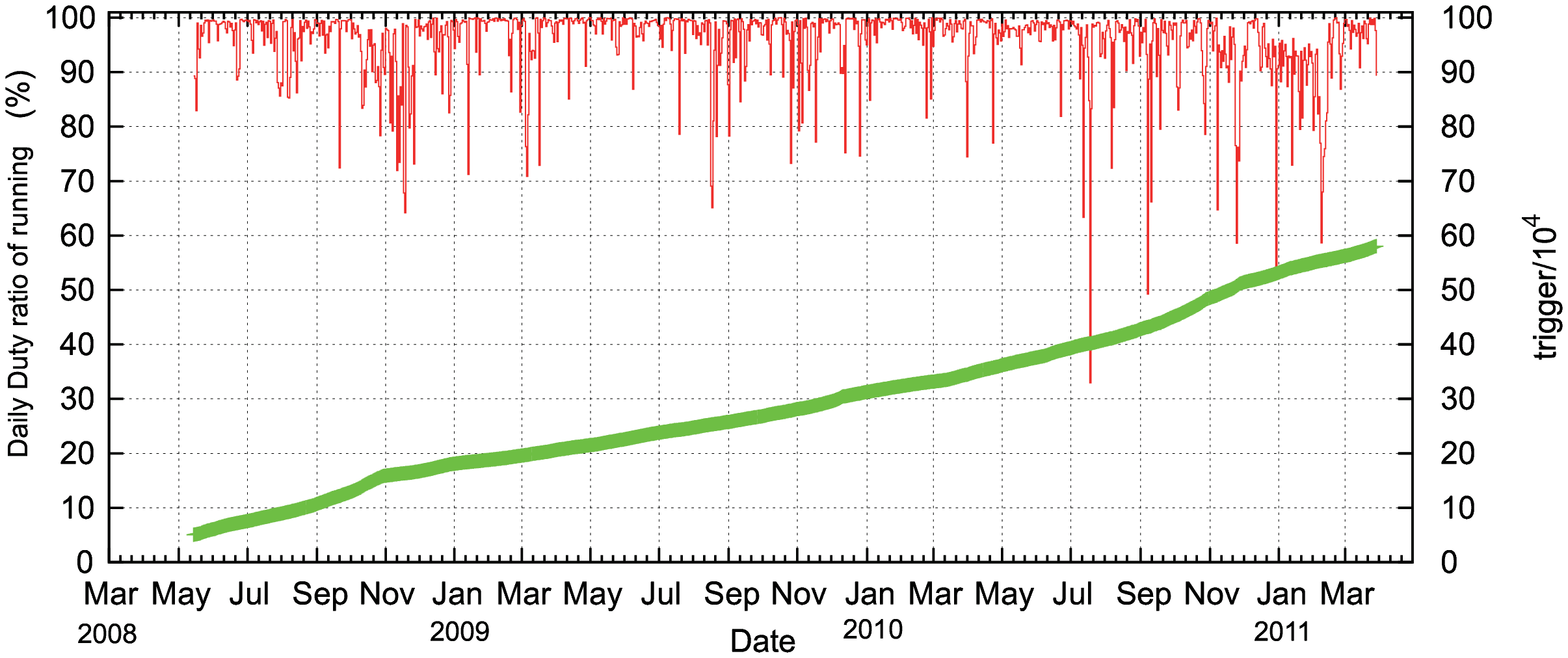}
    \end{center}
    \caption{ Running status of the SD array. Daily duty ratio of running. 
             Evolution of the number of triggered event is also shown using right vertical axis.}
    \label{arraystat}
   \end{figure}
 We showed the running status after three years of SD
operation.
An upgrade of the DAQ system was performed and additional 
deployment was made in November 2008. Including maintenance periods, 
the array has been operating with a 95\% of duty cycle on average.
The variations of detector response and status are recorded in 10 min 
resolution and are well understood. The number of triggers collected as of
March 2011 has reached $5\times10^{5}$.

\section*{Acknowledgments}
The Telescope Array experiment is supported 
by the Japan Society for the Promotion of Science through
Grants-in-Aid for Scientific Research on Specially Promoted Research (21000002) 
``Extreme Phenomena in the Universe Explored by Highest Energy Cosmic Rays'', 
and the Inter-University Research Program of the Institute for Cosmic Ray 
Research;
by the U.S. National Science Foundation awards PHY-0307098, 
PHY-0601915, PHY-0703893, PHY-0758342, and PHY-0848320 (Utah) and 
PHY-0649681 (Rutgers); 
by the National Research Foundation of Korea 
(2006-0050031, 2007-0056005, 2007-0093860, 2010-0011378, 2010-0028071, R32-10130, 2011-0002617);
by the Russian Academy of Sciences, RFBR
grants 10-02-01406a and 11-02-01528a (INR),
IISN project No. 4.4509.10 and 
Belgian Science Policy under IUAP VI/11 (ULB).
The foundations of Dr. Ezekiel R. and Edna Wattis Dumke,
Willard L. Eccles and the George S. and Dolores Dore Eccles
all helped with generous donations. 
The State of Utah supported the project through its Economic Development
Board, and the University of Utah through the 
Office of the Vice President for Research. 
The experimental site became available through the cooperation of the 
Utah School and Institutional Trust Lands Administration (SITLA), 
U.S.~Bureau of Land Management and the U.S.~Air Force. 
We also wish to thank the people and the officials of Millard County,
Utah, for their steadfast and warm support. 
We gratefully acknowledge the contributions from the technical staffs of our
home institutions and the University of Utah Center for High Performance Computing
(CHPC). 



\bibliographystyle{model1-num-names}
\bibliography{<your-bib-database>}

\end{document}